\newif\ifpreprint

\preprintfalse 

\ifpreprint
\documentclass[journal=jpclcd,manuscript=letter]{achemso}
\else
\documentclass[journal=jpclcd,manuscript=letter,layout=twocolumn]{achemso}
\fi

\usepackage[T1]{fontenc} 

\usepackage{amsmath}
\usepackage{newtxtext,newtxmath}

\usepackage{graphicx}
\usepackage{dcolumn}
\usepackage{braket}
\usepackage{multirow}
\usepackage{threeparttable}
\usepackage{xspace}
\usepackage{verbatim}
\usepackage[version=4]{mhchem} 
\usepackage{comment}
\usepackage{color,soul}

\usepackage{mathtools}
\usepackage[dvipsnames]{xcolor}
\usepackage{xspace}
\usepackage{ifthen}

\usepackage{qcircuit}

\usepackage{graphicx,longtable,dcolumn,mhchem}
\usepackage{rotating,color}
\usepackage{lscape}
\usepackage{amsmath}
\usepackage{dsfont}
\usepackage{soul}
\usepackage{physics}
\newcolumntype{d}{D{.}{.}{-1}}
\newcommand{\ctab}[1]{\multicolumn{1}{c}{#1}}
\newcommand{\CheA}{\%{\text{CA}}}
\newcommand{\AccE}{\%{\text{AE}}}
\newcommand{\ie}{\textit{i.e}}
\newcommand{\eg}{\textit{e.g}}
\newcommand{\ra}{\rightarrow}
\newcommand{\pis}{\pi^\star}

\usepackage[colorlinks = true,
            linkcolor = blue,
            urlcolor  = black,
            citecolor = blue,
            anchorcolor = black]{hyperref}

\definecolor{goodorange}{RGB}{225,125,0}
\definecolor{goodgreen}{RGB}{5,130,5}
\definecolor{goodred}{RGB}{220,50,25}
\definecolor{goodblue}{RGB}{30,144,255}

\newcommand{\note}[2]{
\ifthenelse{\equal{#1}{F}}{
\colorbox{goodorange}{\textcolor{white}{\footnotesize \fontfamily{phv}\selectfont #1}}
    \textcolor{goodorange}{{\footnotesize \fontfamily{phv}\selectfont #2}}\xspace
}{}
\ifthenelse{\equal{#1}{R}}{
\colorbox{goodred}{\textcolor{white}{\footnotesize \fontfamily{phv}\selectfont #1}}
    \textcolor{goodred}{{\footnotesize \fontfamily{phv}\selectfont #2}}\xspace
}{}
\ifthenelse{\equal{#1}{N}}{
\colorbox{goodgreen}{\textcolor{white}{\footnotesize \fontfamily{phv}\selectfont #1}}
    \textcolor{goodgreen}{{\footnotesize \fontfamily{phv}\selectfont #2}}\xspace
}{}
\ifthenelse{\equal{#1}{M}}{
\colorbox{goodblue}{\textcolor{white}{\footnotesize \fontfamily{phv}\selectfont #1}}
    \textcolor{goodblue}{{\footnotesize \fontfamily{phv}\selectfont #2}}\xspace
}{}
}

\usepackage{titlesec}

\usepackage[fontsize=11pt]{scrextend}
\titleformat{\section}
{\normalfont\sffamily\bfseries\color{Blue}}
{\thesection.}{0.25em}{\uppercase}

\titleformat{\subsection}[runin]
{\normalfont\sffamily\bfseries}
{\thesubsection}{0.25em}{}[.\;\;]

\titleformat{\suppinfo}
{\normalfont\sffamily\bfseries}
{\thesubsection}{0.25em}{}

\titlespacing*{\section}{0pt}{0.5\baselineskip}{0.01\baselineskip}
\titlespacing*{\subsection}{0pt}{0.125\baselineskip}{0.01\baselineskip}

\author{Pierre-Fran{\c c}ois Loos}
	\email{loos@irsamc.ups-tlse.fr}
	\affiliation[LCPQ, Toulouse]{Laboratoire de Chimie et Physique Quantiques, Universit\'e de Toulouse, CNRS, UPS, France}
\author{Denis Jacquemin}
	\email{Denis.Jacquemin@univ-nantes.fr}
	\affiliation[CEISAM, Nantes]{Laboratoire CEISAM - UMR CNR 6230, Universit\'e de Nantes, 2 Rue de la Houssini\`ere, BP 92208, 44322 Nantes Cedex 3, France}
		
\let\oldmaketitle\maketitle
\let\maketitle\relax

     \title{Is ADC(3) as Accurate as CC3 for Valence and Rydberg Transition Energies?}

\date{\today}

\begin{tocentry}
	\vspace{1cm}
	\includegraphics[width=\textwidth]{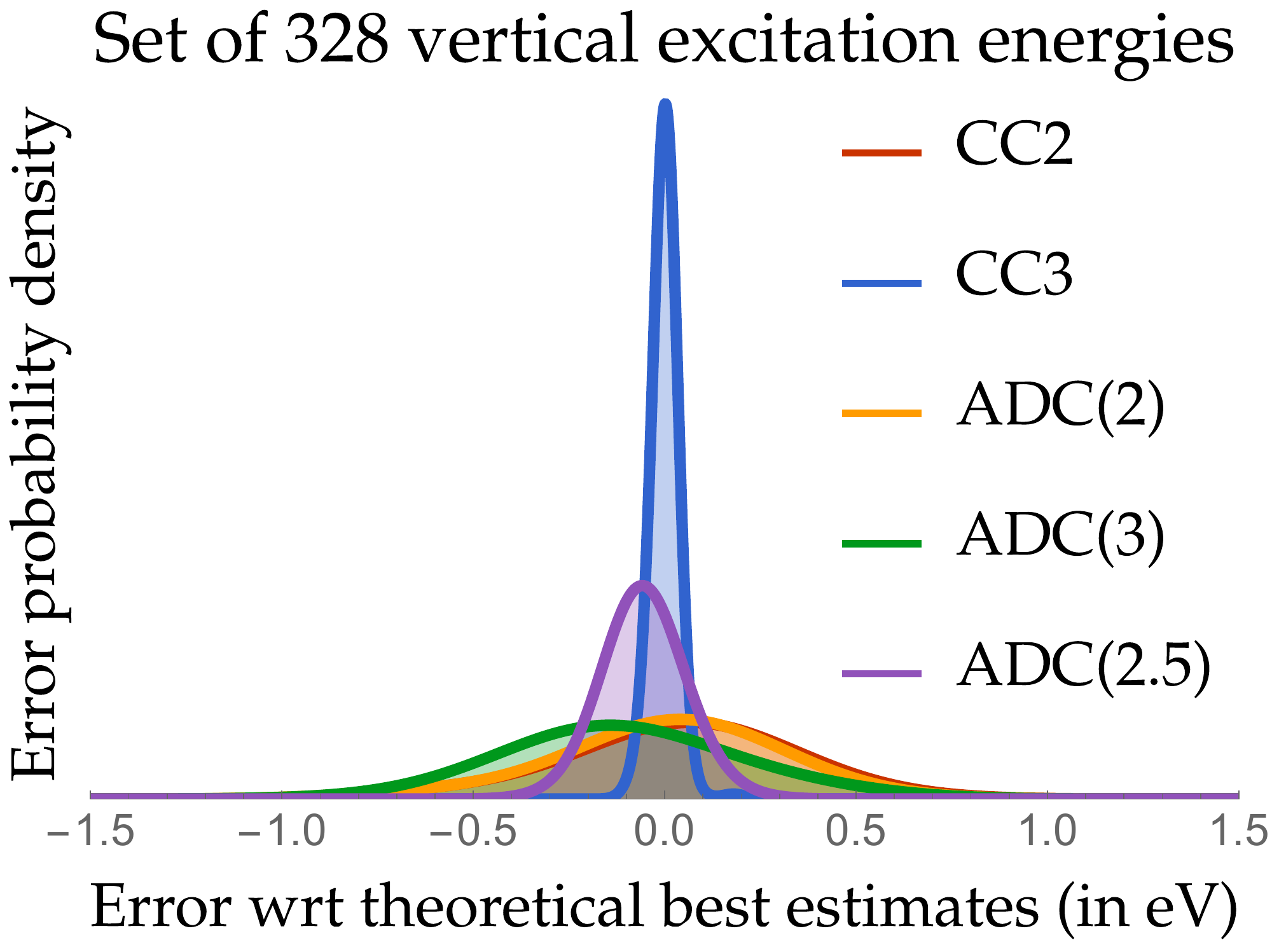}
\end{tocentry}

\begin{document}

\ifpreprint
\else
\twocolumn[
\begin{@twocolumnfalse}
\fi
\oldmaketitle

\begin{abstract}
The search for new \emph{ab initio} models rapidly delivering accurate excited state energies and properties is one of the most active research lines of theoretical chemistry. 
Along with these methodological developments, the performances of known methods are constantly reassessed thanks to the emergence of new benchmark values. In this Letter, 
we show that, in contrast to previous claims, the third-order algebraic diagrammatic construction, ADC(3), does not yield transition energies of the same quality as the
third-order coupled cluster method, CC3.  There is indeed a significant difference in terms of accuracy between the two approaches, as we clearly and unambiguously demonstrate here thanks to extensive comparisons with several
hundreds high-quality vertical transition energies obtained with FCI, CCSDTQ, and CCSDT. Direct comparisons with experimental 0-0 energies of 
small- and medium-size organic molecules support the same conclusion, which holds for both valence and Rydberg transitions, as well as singlet and triplet states. 
In regards of these results, we introduce a composite approach that we named ADC(2.5) which consists in averaging the ADC(2) and ADC(3) excitation energies. 
Although ADC(2.5) does not match the CC3 accuracy, it significantly improves the ADC(3) results, especially for vertical energies. We hope that the present contribution 
will stimulate further developments and, in particular, improvements of the ADC-type methods which have the indisputable advantage of being computationally lighter than their 
equivalent-order CC variants.
\end{abstract}

\ifpreprint
\else
\end{@twocolumnfalse}
]
\fi

\ifpreprint
\else
\small
\fi

\noindent

Electronic excited states (ES) play an important role in many technological applications (photovoltaics, photocatalysis, light-emitting diodes, \ldots), but their characterization from
purely experimental data remains often tedious.  This has stimulated the developments of various density- and wavefunction-based methods allowing to model accurately ES.  Amongst
all these wavefunction approaches, the algebraic diagrammatic construction (ADC), which relies on perturbation theory to access excitation energies and properties, has now become one of
the most popular. \cite{Dre15} The ADC scheme, originally developed by Schirmer and Trofimov, \cite{Sch82,Sch91,Bar95b,Tro97,Tro97b,Sch04d,Sch18} has several advantages over the well-known coupled cluster (CC) 
family of methods, e.g., hermiticity and higher compactness for odd expansion orders. These assets have greatly contributed to the ever growing applications of ADC.  In particular, its second-order 
variant, ADC(2), generally provides valence transition energies as accurate as the one obtained with the second-order CC model, CC2, \cite{Chr95,Hat00} for a smaller computational cost [yet similar $\order*{N^5}$ scaling]. 
\cite{Win13,Har14,Jac15b}  

One of the originality of ADC($n$) lies in its alternative representation, known as intermediate-state representation, of the polarization propagator which poles provides the vertical excitation energies. \cite{Sch82}
These intermediate states are generated by applying a set of creation and annihilation operators to the $n$th-order M{\o}ller-Plesset (MP$n$) ground-state wave function, and are then orthogonalized block-wise according 
to their excitation class. \cite{Sch91} This explains why ADC($n$) is usually presented as ``MP$n$ for excited states'' in the literature. 
One can show that the intermediate states and genuine ES are related by a unitary transformation $\mathbf{X}$, which satisfies the Hermitian eigenvalue problem $\mathbf{M} \mathbf{X} = \mathbf{\Omega} \mathbf{X}$ 
(with $\mathbf{X}^\dag \mathbf{X} = \mathbf{1}$), where $\mathbf{M}$ is the so-called ADC matrix and $\mathbf{\Omega}$ is a diagonal matrix gathering the corresponding excitation energies.
We refer the interested reader to Ref.~\citenum{Dre15} for a non-technical discussion of the general form of the ADC($n$) matrices.

H\"attig pointed out several interesting theoretical connections between ADC(2), CIS(D$_\infty$) and CC2. \cite{Hat05c}
In particular, he showed that ADC(2) is a symmetrized version of CIS(D$_\infty$), and that the only modification required to obtain CIS(D$_\infty$) excitation energies from CC2 is to replace the ground-state CC2 amplitudes by those from MP2.
This idea has been exploited by Dreuw's group to develop the so-called CCD-ADC(2) method where the ADC(2) amplitudes are replaced by those obtained from a coupled cluster doubles (CCD) calculation. \cite{Hod19a,Hod19b}
In addition to improve excitation energies, because CCD-ADC(2) does not rely on perturbation theory anymore, it has been shown to be more robust for molecular dissociation energy curves. \cite{Hod19a}
One of the disadvantages of CC2 compared to ADC(2) is that, due to its non-Hermitian nature, CC2 does not provide a physically correct description of conical intersections between states of the same symmetry, a difficulty absent in ADC(2).

Similarities between the third-order variants, ADC(3) and CC3,  \cite{Chr95b} are likely to exist but, to the best of our knowledge, these potential formal connections have never investigated in the literature.
Nonetheless, it is worth mentioning that CC3, which scales as $\order*{N^7}$, treats the ground state at fourth order of perturbation theory and the 2h-2p block at second order, whereas ADC(3) describes the ground state and 2h-2p block at third and first order of perturbation theory, respectively. \cite{Dre15}
This difference becomes particularly apparent in the calculation of double excitations, for which ADC(3) typically yields inaccurate values. \cite{Loo18a}
However, ADC(3), with its $\order*{N^6}$ computational scaling, has the indisputable advantage of being computationally lighter than CC3, and has a more compact configuration space.

In 2014, Harbach \emph{et al.} \cite{Har14} reported an efficient implementation of ADC(3) and benchmarked its accuracy for transition 
energies using the theoretical best estimates (TBE) of the famous Thiel set \cite{Sch08} as reference.  They concluded that, using the benchmark data available at that time, it was impossible to
determine whether ADC(3) or CC3 was the most more accurate. As ADC(3) enjoys a lower formal computational scaling [$\order*{N^6}$] than CC3 [$\order*{N^7}$], and is generally regarded as the logical path for improvement over ADC(2), this 
finding contributed to enhance the popularity of ADC(3) in the electronic structure community. ADC(3) was subsequently employed to perform theory \textit{vs} experiment comparisons, \cite{Hol15,Boh16,Kni16,Hol17b,Tik19,Sue19,Avi19} 
and to define benchmark values for assessing lower-order methods. \cite{Pla15,Prl16b,Mew16,Aza17b} 

Given, on the one hand, that ADC(3) was advocated as a benchmark method and, on the other hand, the recent availability of high-accuracy reference energies for a large panel of ES,  \cite{Loo18a,Loo19c,Loo20a} 
we believe that the time has come to perform a new performance assessment of this method. To this end, we have first considered our most recent set of TBE/\emph{aug}-cc-pVTZ obtained for vertical 
transition energies in organic compounds encompassing from one to six non-hydrogen atoms.  \cite{Loo18a,Loo20a}  These TBE have been computed at very high levels of theory, \ie, mostly FCI 
(full configuration interaction) for molecules with up to three non-hydrogen atoms, \cite{Loo18a} CCSDTQ for four non-hydrogen atom derivatives, \cite{Loo20a} and CCSDT for compounds containing $5$ or $6$ non-hydrogen atoms.  \cite{Loo20a}  
Note that, for the smallest compounds where the following comparison is actually possible, the mean absolute errors (MAE) obtained with CCSDTQ and CCSDT compared to FCI are trifling ($0.01$ eV and $0.03$ eV, respectively). \cite{Loo18a} 

\begin{figure}
  \includegraphics[width=\linewidth,viewport=2cm 14cm 19cm 27.5cm,clip]{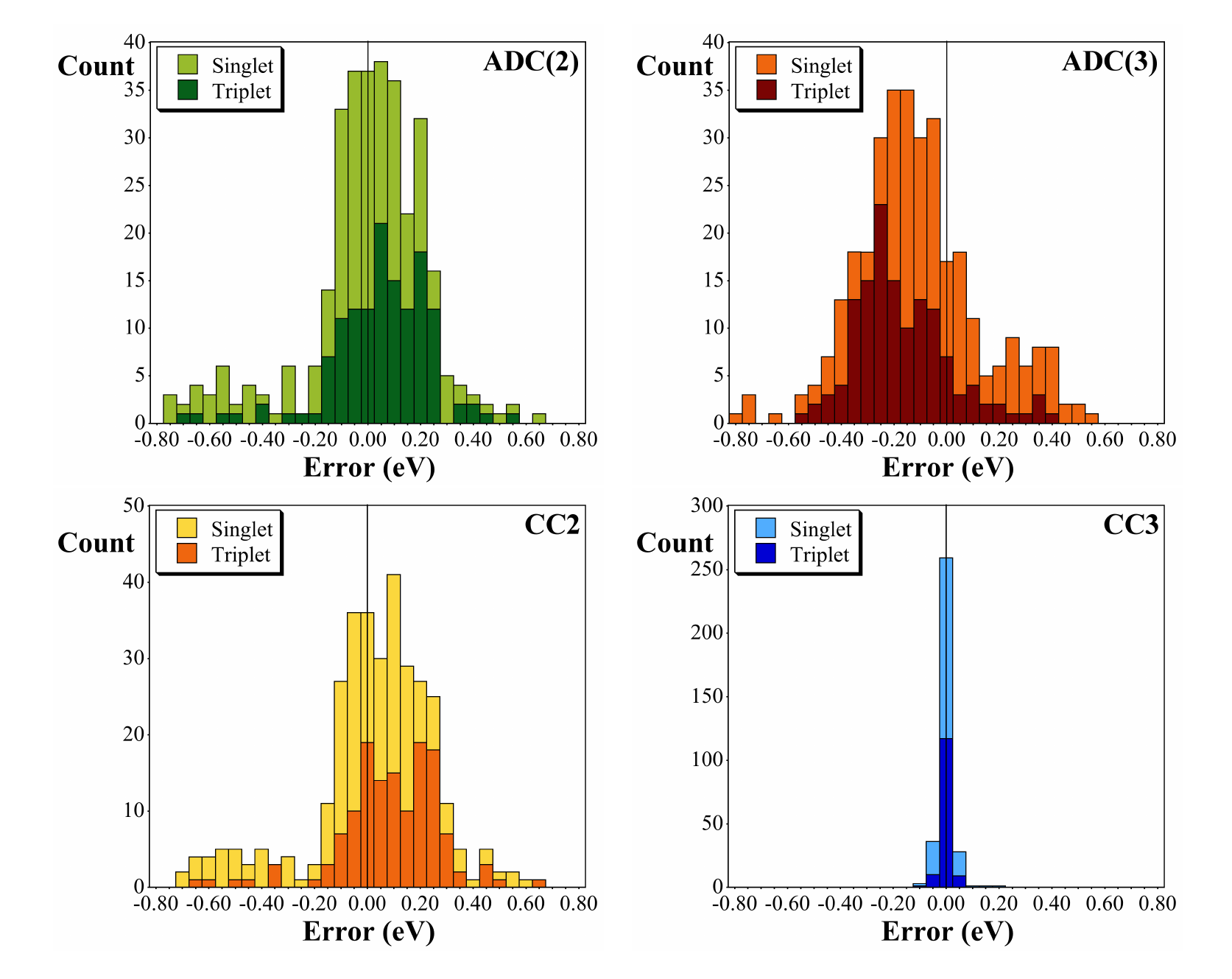}
  \caption{Histograms of the errors (in eV) obtained with ADC(2), ADC(3), CC2, and CC3 taking the TBE/\emph{aug}-cc-pVTZ values of Refs.~\citenum{Loo18a} and \citenum{Loo20a} as reference. 
  ``Count'' refers to the number of transitions in each group.
  The full list of data can be found in the Supporting Information.
  Note the difference of scaling in the vertical axes.}
  \label{Fig-1}
\end{figure}

Table \ref{Table-1} provides a statistical analysis of the performances of the second- and third-order ADC and CC methods, using these TBE as reference. Figure \ref{Fig-1} gives histograms
of the errors for both singlet and triplet states. The full list of data can be found in the Supporting Information.  We consider here a set of $328$ ES, that has been divided into three relatively equivalent 
subsets of $1$--$3$ non-hydrogen atoms ($106$ ES), $4$ non-hydrogen atoms ($89$ ES) and $5$--$6$ non-hydrogen atoms ($134$ ES).  From these data, it is quite clear that CC3 delivers astonishingly accurate transition energies with MAE below or equal to $0.03$ eV for each subset, and no deviation exceeding $\pm 0.20$ eV.  This is inline with several previous benchmarks. \cite{Tro02,Hat05c,Loo18a,Loo18b,Loo19a,Sue19,Loo20a} Again, consistently with 
previous analyses and theoretical considerations (see above), the ADC(2) and CC2 performances are very similar and these second-order methods deliver a global MAE smaller than $0.2$ eV, together with negligible MSE for all subsets. 
This confirms that ADC(2) is indeed a very interesting computational tool thanks to its attractive accuracy/cost ratio. Nevertheless, in par with the above-described conclusions, we found that the performance of ADC(3) is rather 
average with a significant underestimation (MSE of $-0.11$ eV for the full set) and a MAE around $0.20$ eV for each subset.  Overall, ADC(3) underestimates transition energies and provides an 
average deviation of the same order of magnitude as ADC(2) and CC2. Strikingly, the MAE of ADC(3) is basically one order of magnitude larger than the MAE of CC3.

\begin{table}[htp]
\footnotesize
\caption{
Mean signed error (MSE), mean absolute error (MAE), maximal positive error [Max($+$)], and maximal negative error [Max($-$)] with respect to the highly-accurate TBE/\emph{aug}-cc-pVTZ of Refs.~\citenum{Loo18a} and \citenum{Loo20a} 
(see text for details) for various sets of vertical transition energies. All values are in eV. The raw data, which can be found in Table S1 of the Supporting Information, have been obtained with the \emph{aug}-cc-pVTZ basis set and within the frozen-core approximation.}
\label{Table-1}
\begin{tabular}{cldddd}
\hline 
Set 								& Method 		& \ctab{MSE} 	& \ctab{MAE} 	& \ctab{Max($-$)} 	& \ctab{Max($+$)} \\
\hline
All						&	ADC(2)		&0.00	&0.16	&-0.76	&0.64\\
						&	ADC(2.5)		&-0.05	&0.08	&-0.33	&0.24\\
						&	ADC(3)		&-0.11	&0.21	&-0.79	&0.55\\
						&	CC2			&0.02	&0.17	&-0.71	&0.63\\			
						&	CC3			&0.00	&0.02	&-0.09	&0.19\\			
\hline
$1$--$3$ non-H 			&	ADC(2)		&-0.01	&0.21	&-0.76	&0.57 \\
atoms\cite{Loo18a}			&	ADC(2.5)		&-0.08	&0.10	&-0.33	&0.24 \\
						&	ADC(3)		&-0.15	&0.23	&-0.79	&0.39 \\
						&	CC2			& 0.03	&0.22	&-0.71	&0.63 \\
						&	CC3			&-0.01	&0.03	&-0.09	&0.19 \\

$4$ non-H 				&	ADC(2)		&-0.03	&0.18	&-0.73	&0.64\\	
atoms\cite{Loo20a}			&	ADC(2.5)		&-0.07	&0.08	&-0.29	&0.15\\	
						&	ADC(3)		&-0.10	&0.24	&-0.76	&0.49\\	
						&	CC2			& 0.03	&0.20	&-0.68	&0.59\\	
						&	CC3			& 0.00	&0.02	&-0.05	&0.17\\	

$5$--$6$ non-H 			&	ADC(2)		&0.03	&0.11	&-0.48	&0.45\\	
atoms\cite{Loo20a}			&	ADC(2.5)		&-0.02	&0.06	&-0.26	&0.24\\	
						&	ADC(3)		&-0.08	&0.18	&-0.46	&0.55\\	
						&	CC2			&0.01	&0.12	&-0.58	&0.31\\	
						&	CC3			&0.00	&0.01	&-0.03	&0.04\\	
\hline
Valence					&	ADC(2)		&0.07	&0.13&	-0.76&	0.54\\
						&	ADC(2.5)		&-0.05	&0.07&	-0.24	&	0.24\\
						&	ADC(3)		&-0.16	&0.23&	-0.46&	0.50\\
						&	CC2			&0.12	&0.15&	-0.71&	0.50\\
						&	CC3			&0.00	&0.02&	-0.09&	0.19\\
Rydberg					&	ADC(2)		&-0.14	&0.22&	-0.38&	0.64\\
						&	ADC(2.5)		&-0.07	&0.09&	-0.33&	0.24\\
						&	ADC(3)		&-0.01	&0.18&	-0.79&	0.55\\
						&	CC2			&-0.17	&0.21&	-0.41&	0.63\\
						&	CC3			&-0.01	&0.02&	-0.09&	0.17\\				
Singlet					&	ADC(2)		&-0.03	&0.17&	-0.76&	0.64\\
						&	ADC(2.5)		&-0.05	&0.09&	-0.33&	0.24\\
						&	ADC(3)		&-0.07	&0.21&	-0.79&	0.55\\
						&	CC2			&-0.02	&0.18&	-0.71&	0.59\\
						&	CC3			&0.00	&0.02&	-0.09&	0.19\\
Triplet					&	ADC(2)		&0.05	&0.15&	-0.70&	0.57\\
						&	ADC(2.5)		&-0.06	&0.07&	-0.23&	0.19\\
						&	ADC(3)		&-0.17	&0.22&	-0.56&	0.38\\
						&	CC2			&0.09	&0.16&	-0.66&	0.63\\
						&	CC3			&0.00	&0.01&	-0.09&	0.04\\
$n \ra \pis$				&	ADC(2)		&-0.04	&0.09&	-0.38&	0.23\\
						&	ADC(2.5)		&-0.02	&0.06&	-0.23&	0.24\\
						&	ADC(3)		&0.00	&0.14&	-0.32&	0.40\\
						&	CC2			&0.02	&0.08&	-0.25&	0.21\\
						&	CC3			&0.00	&0.01&	-0.05&	0.04\\
$\pi \ra \pis$				&	ADC(2)		&0.14	&0.17&	-0.31&	0.64\\
						&	ADC(2.5)		&-0.07	&0.08&	-0.33&	0.19\\
						&	ADC(3)		&-0.27	&0.29&	-0.79&	0.55\\
						&	CC2			&0.19	&0.21&	-0.41&	0.63\\
						&	CC3			&0.01	&0.02&	-0.09&	0.17\\
\hline																			
\end{tabular}
\end{table}

As can be seen in Table \ref{Table-1}, the ADC(3) MAE obtained for the singlet ($0.21$ eV) and triplet ($0.23$ eV) ES, as well as for valence ($0.23$ eV) and Rydberg ($0.18$ eV) ES are all rather similar.  Interestingly,
ADC(2) exhibits the reverse valence/Rydberg trend with a smaller error for valence transitions ($0.13$ eV) and a larger one for Rydberg ES ($0.22$ eV). It is only for the $n \ra \pis$ transitions ($0.14$ eV) that the ADC(3) MAE 
becomes significantly lower than the usual $0.2$ eV error bar. This success is mitigated by the fact that it is also for the $n \ra \pis$ transitions that ADC(2) and CC2 are the most accurate, 
as both yield MAE smaller than $0.10$ eV for this ES family. On a more optimistic note, one notices that the ADC(3) errors are smallest for the largest compounds gathered in Table \ref{Table-1}. 
This hints that the error might well decrease with system size and become more acceptable for ``real-life'' derivatives. However, a similar trend is observed for both ADC(2) and CC2. 
It is therefore difficult to perform a trustworthy extrapolation to larger systems. 

Finally, as we have found previously, \cite{Loo18a} ADC(3) seems to overcorrect ADC(2).  Therefore, in the spirit of Grimme's and 
Hobza's MP2.5 approach tailored to provide accurate interaction energies, \cite{Pit09} we propose here its excited-state equivalent, ADC(2.5), that simply corresponds to the average between the
ADC(2) and ADC(3) transition energies. Indeed, test numerical experiments have shown that such 50/50 ratio is close to optimal for the present set of transitions. 
 This ADC(2.5) protocol delivers a MSE of $-0.05$ eV and a MAE of $0.08$ eV considering the entire set of transitions.  It is therefore significantly more accurate than ADC(2) or ADC(3) (taken
separately) for practically the same cost as ADC(3). This is well illustrated in Figure S1 of the Supporting Information. This observation might indicate that a renormalized version of ADC(3) could be an interesting alternative to improve its
 overall accuracy, as commonly done for one-electron Green's function methods. \cite{Ced75,Sch83}

\begin{figure}[htp]
  \includegraphics[width=\linewidth,viewport=2cm 14cm 19cm 27.5cm,clip]{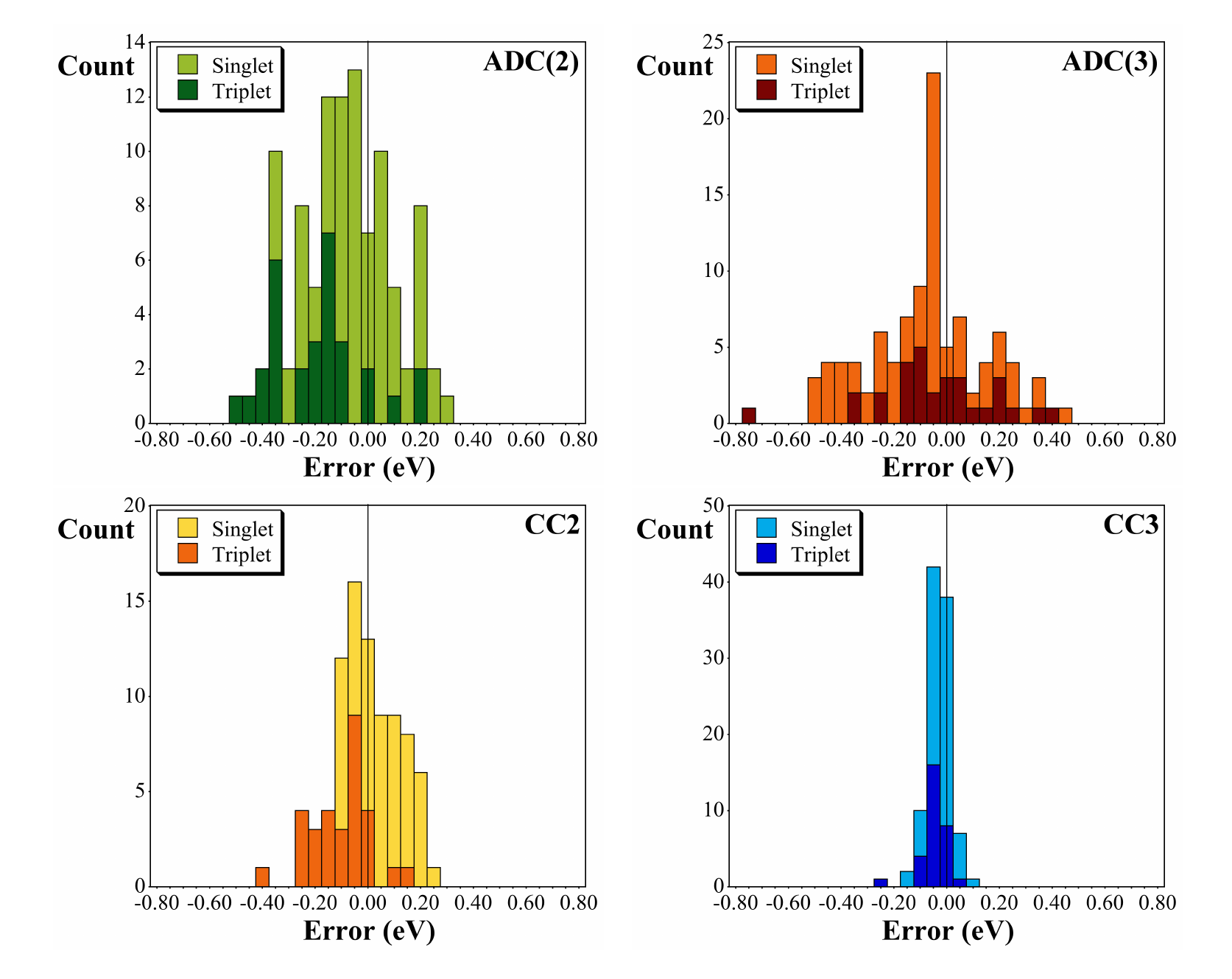}
  \caption{Histograms of the errors (in eV) obtained with ADC(2), ADC(3), CC2, and CC3 taking experimental 0-0 energies as reference. 
  ``Count'' refers to the number of transitions in each group.
  The full list of data can be found in the Supporting Information.
  Note the difference of scaling in the vertical axes.}
  \label{Fig-2}
\end{figure}

Notwithstanding the high accuracy of the vertical excitation energies presented above, CCSDT and CCSDTQ are not error-free. In addition, the previous analysis is limited to compact compounds with a maximum of $6$
non-hydrogen atoms. Therefore, it is worth investigating the correlation between experiment and theoretical observables.  Meaningful theory-experiment comparisons for ES are always challenging but 
the simplest and safest strategy is very likely to be comparing 0-0 energies, an approach that has been used many times before, \eg, see our recent review on the topic. \cite{Loo19b}  Following this strategy, we then consider 
here the (slightly extended) set of compounds defined in Ref.~\citenum{Loo19a}: it encompasses gas-phase measurements for 71 singlet and 30 triplet low-lying transitions.  Note that the typical uncertainty of such 
experimental gas-phase measurements is of the order of $10^{-4}$ eV (or 1 cm$^{-1}$) only. We select here (EOM-)CCSD/\emph{def2}-TZVPP geometries and (TD-)B3LYP/6-31+G(d) vibrational corrections, as it is 
known that the errors in the 0-0 energies are mostly driven by the inaccuracy in the adiabatic energies, rather than the approximate nature of the structures and/or vibrations. \cite{Fur02,Sen11b,Win13,Loo19a} 
Our calculations are again performed with the \emph{aug}-cc-pVTZ basis set, and within the frozen-core approximation. The full list of raw data are given in the Supporting Information.   Statistical data can be found in Table \ref{Table-2}
and Figure \ref{Fig-2}. 

First, considering all 101 cases, we notice that the CC3 adiabatic energies produce chemically accurate 0-0 energies in {59}\%\ of the cases, with errors  almost systematically  smaller than $0.15$ eV.
None of the other approaches can match such a feat. In particular, both ADC(2) and ADC(3) deliver MAE above $0.15$ eV and a chemical accuracy rate smaller than $20\%$. As in the set of vertical transitions
discussed above, ADC(2.5) outperforms ADC(2) and ADC(3), and yields rather small deviations of the same order of magnitude than CC2 (MAE of 0.10 eV).
The fact that CC2 provides more consistent 0-0 energies than ADC(2) while their performances were found similar for vertical
energies might be related to the relatively poorer description of potential energy surfaces with the latter approach. \cite{Bud17}

Turning our attention to the impact of spin symmetry, we note that, although CC3 remains very accurate, we observe a slight decline of its accuracy for triplet ES, a conclusion fitting with our recent study. \cite{Loo19a}
It is also quite clear that ADC(3) has the edge over ADC(2) for triplet ES, whereas the opposite trend is observed for the singlets. Surprisingly, opposite conclusions were drawn for vertical transitions (see above). Despite its tendency to overerestimate (underestimate) singlet (triplet) transition energies (see Figure \ref{Fig-2}), 
CC2 is found to be globally more robust than ADC(2) and ADC(3) for both ES families.  Probably 
more enlightening is the comparison between the results obtained on small (71 molecules with 1--5 non-hydrogen atoms) and medium (30 molecules with $6$--$10$ non-hydrogen atoms) compounds (see Table \ref{Table-2}), the latter set being mostly composed 
of (substituted) six-membered rings. One sees a clear improvement of the ADC(3) performance going from the smaller to the larger molecules, with a MAE of $0.12$ eV and a chemical accuracy rate of $43\%$ for
the latter group. These values are definitively promising. Indeed, although such a MAE value remains {three} times larger than its CC3 analogue, this hints that ADC(3) might become significantly more accurate for larger compounds.
Finally, we wish to recall that these conclusions are made using (EOM-)CCSD geometries and (TD-)DFT harmonic vibrational corrections for all methods. Thus, the overall error is not exclusively 
(though probably predominantly) related to the method selected to compute adiabatic energies. It would be definitely interesting to have access to ground- and excited-state ADC(3) geometries in order to investigate if whether or nor it yields an improvement of the ADC(3) performance. \cite{Reh19}

\begin{table}[htp]
\footnotesize
\caption{Mean signed error (MSE) and mean absolute error (MAE), as well as percentage of chemical accuracy (\CheA, absolute error below 0.043 eV) and acceptable error (\AccE, absolute error below 0.150 eV) with respect 
to experimental 0-0 energies for the (71) singlet and (30) triplet sets of 0-0 energies from Ref.~\citenum{Loo19a}.  All values are in eV and have been obtained with the \emph{aug}-cc-pVTZ basis set and within the frozen-core approximation 
using (EOM-)CCSD/\emph{def2}-TZVPP geometries and (TD-)B3LYP/6-31+G* vibrational corrections. The full list of data can be found in the Supporting Information.
}
\label{Table-2}
\begin{tabular}{clddcc}
\hline 
Set 				& Method 	& \ctab{MSE} 	& \ctab{MAE} 	& \ctab{\CheA} &  \ctab{\AccE}   \\
\hline
All					&	ADC(2)	& -0.09	&0.16	&18	&52\\
					&	ADC(2.5)	& -0.08	&0.10	&24	&78\\
					&	ADC(3)	& -0.07	&0.18	&19	&50\\
					&	CC2		& 0.00	&0.10	&31	&75\\
					&	CC3		&-0.03	&0.04	&59	&98 \\
\hline
$1$--$5$ non-H 		&	ADC(2)	& -0.10	&0.16	&15	&55\\
atoms				&	ADC(2.5)	& -0.11	&0.11	&24	&72\\
					&	ADC(3)	& -0.13	&0.21	&8	&41\\
					&	CC2		& 0.01	&0.09	&31	&82\\
					&	CC3		&-0.03	&0.05	&62	&97 \\
$6$--$10$ non-H 		&	ADC(2)	& -0.07	&0.17	&23	&47\\
atoms				&	ADC(2.5)	& -0.01	&0.06	&23	&97\\
					&	ADC(3)	& 0.05	&0.12	&43	&70\\
					&	CC2		& -0.11	&0.11	&30	&60\\
					&	CC3		& -0.03	&0.04	&53	&100\\
\hline
Singlet				&	ADC(2)	& -0.05	& 0.13	&	23	& 62 \\
					&	ADC(2.5)	& -0.07	&0.09	&	31	&76	\\
					&	ADC(3)	& -0.09	& 0.19	&	18	& 48 \\
					&	CC2		& +0.05	& 0.09	&	34 	& 80 \\
					&	CC3		& -0.03	& 0.04	&	63	& 99 \\
Triplet				&	ADC(2) 	& -0.20	& 0.23	&	7	& 30 \\
					&	ADC(2.5)& -0.12	&0.12	&	7	& 87	\\
					&	ADC(3) 	& -0.04	& 0.17	& 	20	& 53 \\
					&	CC2		& -0.11	& 0.12	&	23	& 63 \\
					&	CC3		& -0.05	& 0.05	&	50	& 97 \\
\hline																			
\end{tabular}
\end{table}

At this stage, it seems natural to wonder why the conclusions of the 2014 ADC(3) assessment \cite{Har14} based on Thiel's set differ significantly from ours although the nature of the molecules belonging to the 
two sets are relatively similar. To understand this discrepancy, let us reexamine the data of Ref.~\citenum{Har14}. In this work, Thiel's original TBE (denoted as TBE-1), \cite{Sch08} mostly based on CASPT2/TZVP 
but also incorporating some CC3/TZVP (as well as other values), were used as reference rather than Thiel's most recent set of TBE (denoted as TBE-2), \cite{Sil10c} which are mostly basis set corrected CC3/\emph{aug}-cc-pVTZ 
values.  In addition,  given the knowledge at that time, the authors of Ref.~\citenum{Har14} logically decided that to consider only the non-CC3 TBE  values in their comparison of the relative accuracy of ADC(3) and CC3, 
which is a very reasonable point. Considering the subset of TBE-1 based on CASTP2 (\ie, excluding the CC3 values from TBE-1), 
Ref.~\citenum{Har14} reports, for the singlet states, a MSE (MAE) of $+0.23$ ($0.24$) eV for CC3. This value has to be compared with a MSE (MAE) of $+0.12$ ($0.24$) eV for ADC(3) where the reference was 
taken as the entire TBE-1 set. \cite{Har14}  Similarly, for the 19 triplet excitation energies of the TBE-1 set not based on CC3, the MSE is $+0.12$ eV with CC3 and $-0.10$ eV with ADC(3). \cite{Har14}
The direct comparison of ADC(3) and CC3 is also instructive. By considering now CC3 as reference, the MSE (MAE) of ADC(3) reported in Ref.~\citenum{Har14} are $-0.20$ ($0.29$) eV for the singlets and 
$-0.22$ ($0.25$) eV for the triplets. \cite{Har14}  These numbers are consistent with the findings of the present Letter, and show that ADC(3) significantly underestimates both families of transitions.  
We can then conclude that the bias in this earlier ADC(3) assessment \cite{Har14} was likely due to the CASPT2 reference values. Indeed, as clearly demonstrated in a recent series of papers 
\cite{Loo18a,Loo18b,Loo19a,Sue19,Loo20a}, CC3 is a very robust method which generally delivers chemically accurate excitation energies, while CASPT2 has a clear tendency of underestimating transition energies. \cite{Loo20a} 

In this context, we also wish to point out that an early ADC(3) \emph{vs} 
FCI benchmark performed for a series of small molecules (\ce{H2O}, \ce{HF}, \ce{N2}, \ce{Ne}, \ce{CH2}, and \ce{BH}), \cite{Tro02} concluded that \emph{``the mean absolute error, 
as calibrated versus the FCI results for 41 singlet and triplet transitions, has been found to be smaller than $0.2$ eV''} (more precisely the MAE is equal to $0.18$ eV for the first four compounds) and 
that \emph{``the quality of the results [...] does not match the impressive accuracy of the CC3 computations''}. The present results confirm these two earlier assertions.  

An additional aspect to take into account is that previous comparisons between ADC(3) transition energies and experimental $\lambda_{\mathrm{max}}$ values were often performed in the vertical approximation, \cite{Kni16,Mew18} 
which means that the geometry relaxation and vibronic effects were neglected, which is often done, as such vibronic corrections are computationally expensive.  However, as shown in several works, \cite{Die04b,Goe10a,Sen11b,Jac12d,Fan14b,San16b} 
this approximation implies a significant bias, because  the blueshift between the experimental 0-0 energy and the $\lambda_{\mathrm{max}}$ value is typically smaller than the blueshift between the computed 0-0 and vertical energies. 
As a consequence, applying the vertical approximation favors methods delivering smaller transition energies. 

As an example, the $Q$-band of \ce{Mg}-porphyrin was studied at various levels of theory 
including ADC(3) in 2018. \cite{Mew18}  The first experimental maxima appears at $2.07$ eV, \cite{Edw71} a value smaller than the ADC(2), CCSD, and TD-DFT vertical transitions (which are found in the $2.27$--$2.43$ eV range) 
as it should. \cite{Mew18} In contrast, the ADC(3) vertical value of $2.00$ eV, is the closest from experiment but presents the incorrect error sign and would likely be significantly too redshifted if 
vibronic corrections were accounted for. Indeed, according to Durbeej, \cite{Fan14b} the CC2 difference between vertical and 0-0 energies is $-0.05$ eV in the (free-base) porphyrin. This brings the 
ADC(3) estimate to $-0.12$ eV compared to experiment and improves the agreement for the other approaches. Again, both the error sign and its magnitude are quite coherent with the present estimates.  
Using the same procedure, ADC(2.5) would give a 0-0 energy of $2.11$ eV, in superb agreement with experiment.

In the same work, \cite{Mew18} an ADC(3) value of $4.65$ eV is reported for the lowest $B_u$ state of \emph{trans}-octatetraene, a bright ES with a dominant single-excitation character. 
\cite{Mew18}  This value is significantly lower than Thiel's CC3 value of $4.84$ eV, \cite{Sil10c} although the latter was obtained on a MP2 geometry that slightly underestimates the bond length alternation, 
whereas the ADC(3) estimate relies on a more accurate CCSD(T) structure.  The measured gas-phase 0-0 energy for this transition is $4.41$ eV, \cite{Leo84} and the estimated difference between 
vertical and 0-0 energies is $-0.45$ eV at the TD-BHHLYP level, \cite{Die04b} and $-0.36$ eV at the CC2 level, \cite{Fan14b} again hinting that the ADC(3) value is in fact slightly too low by a magnitude of $-0.12$ eV 
if one naively applies the CC2 correction (determined on a CC2 geometry). In this case, ADC(2.5) would only slightly reduced the error to $-0.10$ eV. 

Of course, these two comparisons remain very qualitative and one would greatly benefit from  ADC(3) 0-0 energies which, to the best of our knowledge, are not available to date for these compounds. 

In this Letter, we have provided what we believe are compelling evidences that the transition energies computed with ADC(3) in organic compounds are significantly less accurate than their CC3 counterparts. 
This statement is based on i) extensive comparisons with both vertical energies determined with higher levels of theory (CCSDT, CCSDTQ, and FCI), and ii) accurate 0-0 energies measured in gas phase for small-
and medium-size compounds. This conclusion apparently holds almost irrespectively of the nature of the transition, provided that the ES does not exhibit a dominant double excitation character.  Of course, given
that the ADC(3) error for 0-0 energies has a clear tendency to significantly drop for the largest compounds considered here (\ie, substituted six-membered rings), one could rightfully speculate that ADC(3) 
would become more accurate for even larger compounds, a claim that we cannot honestly verify at this stage.  Besides, ADC(3) might also deliver accurate ES properties (such as geometries, transition and total 
dipoles, oscillator strengths, two-photon cross-sections, etc). Indeed, these properties are treated at third order of perturbation theory by both ADC(3) and CC3. We believe that comparisons between CC3 and ADC(3) properties is a particular
 point that needs to be further investigated in the future.

\section*{Computational details}
For the set of vertical transition energies, the CC3/\emph{aug}-cc-pVTZ geometries of Refs.~\citenum{Loo18a} and \citenum{Loo20a} have been selected because the TBE have been obtained on the
very same structures. The GS and ES structures used in the 0-0 calculations have been obtained at the (EOM-)CCSD/\emph{def2}-TZVPP level and are provided in the Supporting Information of Ref.~\citenum{Loo19a}.
The zero-point vibrational energies used to compute the 0-0 energies have been (mostly) obtained at the (TD-)B3LYP/6-31+G(d) level and are all listed in the Supporting Information of Ref.~\citenum{Loo19a}. The CC and ADC calculations 
have been performed with DALTON \cite{dalton} and Q-CHEM, \cite{Sha15} respectively, with the \emph{aug}-cc-pVTZ basis set. The ADC calculations have been performed within the RI approximation. Test calculations
have shown that this approximation implies only trifling changes in the transition energies ($< 0.01$ eV). We refer the readers to our previous works \cite{Loo18a,Loo19a} for additional details.

\section*{Acknowledgements}
PFL thanks the \textit{Centre National de la Recherche Scientifique} for funding.
This research used resources of i) the GENCI-CINES/IDRIS; ii) CCIPL (\emph{Centre de Calcul Intensif des Pays de Loire}); iii) a local Troy cluster and iv) HPC resources from ArronaxPlus 
(grant ANR-11-EQPX-0004 funded by the French National Agency for Research). 

\section*{Supporting Information Available}
Full list of transition energies for vertical and 0-0 energies.

\bibliography{biblio-new}

\end{document}